# Prototyping the use of Large Language Models (LLMs) for adult learning content creation at scale

Daniel Leiker [1,2], Sara Finnigan [2], Ashley Ricker Gyllen [2,3] and Mutlu Cukurova [1]

[1] UCL Knowledge Lab, London WC1N 3QS, UK
[2] EIT InnoEnergy, Eindhoven, NL
[3] Metropolitan State University, Denver CO 80204, USA

**Abstract**

As Large Language Models (LLMs) and other forms of Generative AI permeate various aspects of our lives, their application for learning and education has provided opportunities and challenges. This paper presents an investigation into the use of LLMs in asynchronous course creation, particularly within the context of adult learning, training and upskilling. We developed a course prototype leveraging an LLM, implementing a robust human-in-the-loop process to ensure the accuracy and clarity of the generated content. Our research questions focus on the feasibility of LLMs to produce high-quality adult learning content with reduced human involvement. Initial findings indicate that taking this approach can indeed facilitate faster content creation without compromising on accuracy or clarity, marking a promising advancement in the field of Generative AI for education. Despite some limitations, the study underscores the potential of LLMs to transform the landscape of learning and education, necessitating further research and nuanced discussions about their strategic and ethical use in learning design.

**Keywords**

Large Language Models, Generative AI in Education, AI-Generated Learning Content

## 1. Introduction

In the past year, Large Language Models (LLMs), a specific type of Generative AI focused on generating human-like text, have been introduced into mainstream and professional use. This has sparked debate around the opportunities and challenges that come with leveraging these tools for learning and education [e.g., 1, 2]. Already, LLMs are being applied in formal educational contexts for children and adults alike; in AI-powered chatbots [3], for grading student's work [4, 5], and in generating study materials and assessments [6, 7]. However, less has been discussed in the literature with respect to the application of LLMs in informal adult learning contexts, despite the growing global demand for educational content designed for online training programs and employee upskilling. At first glance, LLMs seem to be an ideal tool for creating novel learning content, given their ability to access and synthesize a wide range of human knowledge into specific outputs through careful inputs to the model or prompts [i.e., prompt engineering, cf 8]. However, concerns about the accuracy and reliability of the generated outputs have led to hesitation among many learning professionals and institutions in endorsing their widespread use. In response to these concerns, this paper presents an investigation into the use of LLMs in asynchronous course creation within the context of adult learning and reskilling, using a human-in-the-loop approach.

---

A common challenge in creating high-quality learning content for online instructional platforms is the limited availability of individuals who can provide crucial subject matter expertise. The application of LLMs in learning design has the potential to address this challenge and transform the process by which educational experiences are created, expediting instructional design and decreasing the amount of time needed from subject matter experts. Regardless of whether the content development process can be accelerated by leveraging LLMs, what remains critical is the implementation of sound instructional design rooted in learning science and evidence-based practices [9, 10, 11]. The importance of well-defined objective design, robust assessment strategies, and tight instructional alignment remain paramount. These elements are vital to building online learning experiences that are not only efficient and effective, but also successful in achieving desired learning outcomes [12, 13].

This study explores the potential of leveraging LLMs to enhance asynchronous course creation through a robust human-in-the-loop process. Examining the extent to which learning designers can guide LLMs, with strategic and efficient input from subject matter experts, to produce educational content of sufficient accuracy and clarity. We hypothesize that LLMs, coupled with rigorous instructional design practices, can accelerate the course development process while retaining expected quality standards. In this paper, we address the following research question:

*Does course content created with LLMs using a human-in-the-loop process differ significantly in accuracy or clarity from course content created by human subject matter experts?*

## 2. Methodology
## 2.1. Content Creation Process

To create the experimental course content, an OpenAI LLM was used along with prompt engineering guidance developed by the content creation team as part of a robust human-in-the-loop process. This team consisted of a lead learning designer, an architect for AI-driven learning experiences, and an instructional designer. The team was also minimally supported by a subject matter expert. The content creation process began using a backwards design instructional approach [14], developing a list of course topics with the support of the subject matter expert. The lead learning designer applied prompt engineering guidance to design a course outline with course objectives using the LLM. The instructional designer then used the LLM, with prompt engineering guidance, to generate the text-based content elements, following a course blueprint template designed to define requirements for the scope and sequence of activities within the course. Applying this process, the content creation team was able to design and develop a multi-lesson, multimedia-driven course, including all the course content elements (i.e., course outline, course text, video scripts, interactive activities, and assessments).

After the design and development of the course was completed, the course blueprint was rated and reviewed by electrical engineering subject matter experts (described below). Based on these reviews, edits were made to the blueprint, and it was sent to a small team of developers contracted to build the final course using Storyline 360 for building the course shell and content elements and Synthesia (a text-to-video Generative AI tool) to create the videos. The end result was a 1.5 hour, 3-lesson course titled Basic Concepts of Electrical Systems targeting adult learners in the renewable energy industry. We estimate the design and development process to have taken 22.5 hours (~15 hours per every hour of intended learner effort time). This represents a substantial decrease from the typical time to design a course by up to 25 times.

## 2.2. Content Rating Process

Five (5) subject matter experts were invited to review and rate the course design at the blueprint stage for accuracy and clarity. While this was a convenience sample, the rater population was defined as individuals with electrical engineering (EE) knowledge and experience. The five raters that participated were all individuals with a working knowledge of the subject matter, formal education in

EE (2 Bachelor's degrees; 2 Masters degrees, 1 PhD), and experience working as EEs (Mean Experience = 10.2 years).

A rating sheet was created with clear instructions to rate 20 specific course content elements and sections (e.g., course text, video scripts, interactive activities, and assessments) for both accuracy and clarity on a 3-point Likert scale. Using this rating sheet, each rater completed two reviews, one of the experimental course content described above and one of the control course content. The control course content consisted of a course blueprint from a course titled The Energy System: Present and Future. This course was similar in length and complexity to the experimental course and was also designed to target adult learners in the renewable energy industry. However, the control course was created using traditional design and development methods (i.e., a collaboration between learning designers and faculty members with subject expertise). Ratings for both courses were independent, as there was no communication or consensus between raters. Additionally, raters were blind to the course content creation method (i.e., they were not aware that one of the courses was generated by artificial intelligence until after the reviews were completed).

## 3. Results

To evaluate our research question, ratings of accuracy and clarity were evaluated separately. For each variable, average ratings across the 20 elements were calculated within each rater once for the experimental course content and again for the control course content. The weighted averages across all 5 raters are presented in Figure 1. Rater averages were determined to be normal and free of outliers, therefore comparisons between the two courses were conducted using paired-sample t-tests. To examine the extent to which our 5 raters agreed with each other in their assessment of the course content, inter-rater reliability analyses were conducted using Kendall's coefficient of concordance ($W$). This statistic was chosen because the ratings were ordinal and independent. All analyses were completed using relevant packages in R.

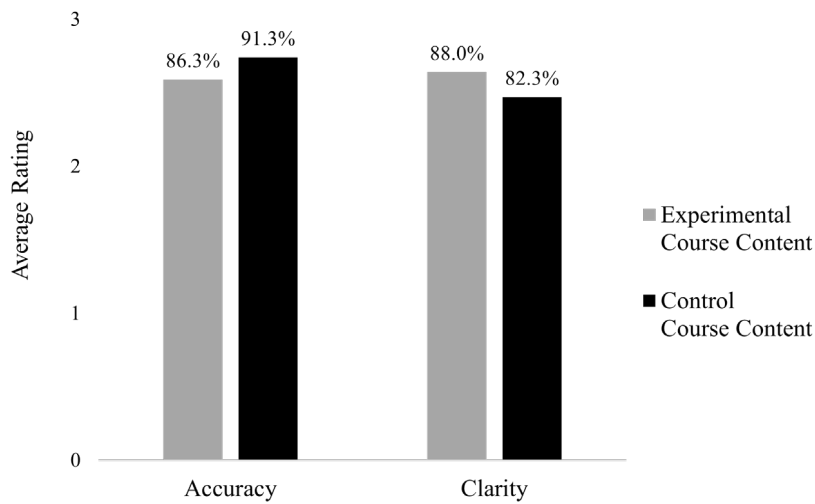

**Figure 1**: Average Ratings for Accuracy and Clarity of Experimental Course Content (grey) and Control Course Content (black)

### 3.1. Content Accuracy

For the experimental course, expert raters gave the content an average accuracy rating of 86.3%. Accuracy of the control course content was rated as higher than (91.3%) but not significantly different from the experimental course condition ($p = .16$). Inter-rater reliability analyses demonstrated that the 5 expert raters had a high level of agreement among their independent assessments of accuracy for both the experimental course content ($W = .93$) and the control course content ($W = .96$).

## 3.2. Content Clarity

Expert raters gave the experimental course content an average clarity rating of 88.0%. In contrast to accuracy, clarity of the control course content was rated as lower than (82.3%) although still not significantly different from the experimental course content ($p = .09$). Inter-rater reliability analyses again demonstrated a high level of agreement among the expert raters' independent assessments of clarity for both the experimental course content ($W = .91$) and the control course content ($W = .92$).

## 4. Discussion

The findings of this project reveal that Large Language Models (LLMs), combined with a robust human-in-the-loop process, can be effectively leveraged to design and develop high quality, comprehensive, multimedia-driven courses. Expert reviewers evaluated a course designed and developed using this type of approach, comparing it with a course using a traditional design and development process, and the result was a near parity between the two. Additionally, the accuracy and clarity of the courses were interpreted almost identically across all five of the expert raters. This suggests that Generative AI, in the form of LLMs, can indeed be a viable tool for creating accurate and clear educational content. Although the findings were not significant, it is notable that the content created with the support of LLMs slightly outperformed the traditionally created content in terms of clarity, though it was slightly less accurate. These findings substantiate the initial thesis that a large language model could, with proper guidance, generate accurate and clear adult learning courses within a relatively short period of time.

As the demand grows for online training programs and employee upskilling, the need for accessible, clear, and accurate educational materials grows. Our findings indicate that the use of LLMs to generate online, interactive educational content offers a promising solution to this challenge. Further, these findings suggest that by utilizing a well-designed human-in-the-loop process, content created using LLMs can match traditionally created educational content in both accuracy and clarity. Importantly, our definition of "well-designed" includes giving human input to the model from both subject matter experts as well as learning designers well versed in learning science and evidence-based practices [10, 11]. Similarly, we have advocated for the pairing of Generative AI and sound learning design principles in a recent study investigating the use of AI-powered text-to-video tools to facilitate effective learning videos [15]. These two works build off each other, and taken together, stand as an exciting advancement within the field of Generative AI for education, in starting to address the lack of literature examining the application of LLMs in informal adult learning contexts.

One limitation of this study exists with respect to scope. A deeper dive into the effectiveness of our prompt engineering guidance is needed, given the importance of effective prompts when utilizing LLMs [8]. Indeed, effective prompt engineering is an integral part of the course creation process and of which the human-in-the-loop process is most critically and innately dependent. Better understanding of the limitations to our approach, and best practices in prompt engineering in general [16] may be a key factor at this stage to ensure engaging and accurate learning experiences.

## 5. Conclusion

This study validates the potential of Large Language Models in accelerating the creation of educational content without compromising accuracy or clarity. Comparisons between course content created using LLMs and traditional, human-created content underline the promise of AI in reshaping the landscape of learning and education, particularly in adult learning, training, and upskilling. However, the incorporation of LLMs introduces new roles for learning designers and necessitates robust instructional design practices. Therefore, as we harness the benefits of AI in education, ethical considerations, quality control measures, and strategic integration become imperative for realizing its full potential.

## 6. Acknowledgements

We gratefully acknowledge EIT InnoEnergy Skills Institute for their support of the research, and for allowing us to examine their traditionally generated course as our control. We would also like to thank Blue Carrot for the excellent production work of their development team in building the final course.

## 7. References

[1] Cooper, G. (2023). Examining science education in ChatGPT: An exploratory study of generative artificial intelligence. Journal of Science Education and Technology, 1-9. doi:10.1007/s10956-023-10039-y.

[2] Kasneci, E., Sessler, K., Küchemann, S., Bannert, M., Dementieva, D., Fischer, F., ... & Kasneci, G. (2023). ChatGPT for good? On opportunities and challenges of large language models for education. Learning and Individual Differences, 103, 102274. doi:10.1016/j.lindif.2023.102274.

[3] Khan, S. (2023, March 14). Harnessing GPT-4 so that all students benefit. A nonprofit approach for equal access. Khan Academy. URL: https://blog.khanacademy.org/harnessing-ai-so-that-all-students-benefit-a-nonprofit-approach-for-equal-access/

[4] Moore, S., Nguyen, H. A., & Stamper, J. (2020). Evaluating crowdsourcing and topic modeling in generating knowledge components from explanations. In Artificial Intelligence in Education: 21st International Conference, AIED 2020, Ifrane, Morocco, July 6–10, 2020, Proceedings, Part I 21 (pp. 398-410). Springer International Publishing.. doi:10.1007/978-3-030-52237-7_32.

[5] Bauer, E., Greisel, M., Kuznetsov, I., Berndt, M., Kollar, I., Dresel, M., ... & Fischer, F. (2023). Using natural language processing to support peer-feedback in the age of artificial intelligence: A cross-disciplinary framework and a research agenda. British Journal of Educational Technology. doi:10.1111/bjet.13336.

[6] Dijkstra, R., Genç, Z., Kayal, S., & Kamps, J. (2022). Reading Comprehension Quiz Generation using Generative Pre-trained Transformers.

[7] Gabajiwala, E., Mehta, P., Singh, R., & Koshy, R. (2022, November). Quiz Maker: Automatic Quiz Generation from Text Using NLP. In Futuristic Trends in Networks and Computing Technologies: Select Proceedings of Fourth International Conference on FTNCT 2021 (pp. 523-533). Singapore: Springer Nature Singapore. doi:10.1007/978-981-19-5037-7_37.

[8] Liu, V., & Chilton, L. B. (2022, April). Design guidelines for prompt engineering text-to-image generative models. In Proceedings of the 2022 CHI Conference on Human Factors in Computing Systems (pp. 1-23). doi:10.1145/3491102.3501825.

[9] Sawyer, R. K. (Ed.). (2005). The Cambridge handbook of the learning sciences. Cambridge University Press. doi:10.1017/cbo9780511816833.

[10] Mayer, R. E. (2019). Thirty years of research on online learning. Applied Cognitive Psychology, 33(2), 152-159. doi:10.1002/acp.3482.

[11] Goodell, J., & Kolodner, J. (Eds.). (2022). Learning Engineering Toolkit: Evidence-Based Practices from the Learning Sciences, Instructional Design, and Beyond. Taylor & Francis. doi: 10.4324/9781003276579.

[12] Osborne, R. J., & Wittrock, M. C. (1983). Learning science: A generative process. Science education, 67(4), 489-508. doi:10.1002/sce.3730670406.

[13] Clark, R. C., & Mayer, R. E. (2016). E-learning and the science of instruction: Proven guidelines for consumers and designers of multimedia learning. John Wiley & Sons. doi:10.1002/9781119239086.

[14] Wiggins, G. & McTighe, J. (2005). Understanding by design (2nd edition). ASCD.

[15] Leiker, D., Gyllen, A. R., Eldesouky, I., & Cukurova, M. (2023). Generative AI for learning: Investigating the potential of synthetic learning videos. arXiv preprint arXiv:2304.03784.

[16] White, J., Fu, Q., Hays, S., Sandborn, M., Olea, C., Gilbert, H., ... & Schmidt, D. C. (2023). A prompt pattern catalog to enhance prompt engineering with chatgpt. arXiv preprint arXiv:2302.11382.